\documentclass[a4paper,12pt]{article}
\usepackage{epsfig}
\usepackage{amssymb}
\usepackage{amsfonts}
\usepackage{amsmath}
\usepackage{euscript}
\usepackage{verbatim}
\usepackage{latexsym}
\usepackage{graphicx}

\usepackage{tikz}
\usetikzlibrary{shapes.geometric, arrows,patterns,snakes}
\tikzstyle{ellip} = [ellipse, minimum width=3cm, minimum height=1cm,text centered, draw=black]

\newskip\humongous \humongous=0pt plus 1000pt minus 1000pt

\newif\ifdtup

\catcode`\@=11

\@addtoreset{equation}{section}

\def\@normalsize{\@setsize\normalsize{15pt}\xiipt\@xiipt
\abovedisplayskip 14pt plus3pt minus3pt%
\belowdisplayskip \abovedisplayskip
\abovedisplayshortskip \z@ plus3pt%
\belowdisplayshortskip 7pt plus3.5pt minus0pt}

\def\small{\@setsize\small{13.6pt}\xipt\@xipt
\abovedisplayskip 13pt plus3pt minus3pt%
\belowdisplayskip \abovedisplayskip
\abovedisplayshortskip \z@ plus3pt%
\belowdisplayshortskip 7pt plus3.5pt minus0pt
\def\@listi{\parsep 4.5pt plus 2pt minus 1pt
     \itemsep \parsep
     \topsep 9pt plus 3pt minus 3pt}}

\relax

\catcode`@=12

\catcode`\@=11

\def\section{\@startsection{section}{1}{\z@}{3.5ex plus 1ex minus
   .2ex}{2.3ex plus .2ex}{\large\bf}}

\def\SymBoxes#1#2#3#4{\newdimen\un@t \un@t#3%
\raisebox{#1}{\rule{#2\un@t}{#4}\hskip-#2\un@t
\@tempdimb\un@t \advance\@tempdimb by-#4\@tempcntb#2\relax%
\@whilenum{\@tempcntb>0}\do{
\rule{#4}{\un@t}\hskip\@tempdimb \advance\@tempcntb by\m@ne}%
\hskip-#2\un@t \rule[\un@t]{#2\un@t}{#4}%
\rule[\un@t]{#4}{#4}\hskip-#4
\rule{#4}{\un@t}}\hskip-#4}                

\begin{document}

\newcommand{\beq}{\begin{equation}}
\newcommand{\eeq}{\end{equation}}
\newcommand{\bea}{\begin{eqnarray}}
\newcommand{\eea}{\end{eqnarray}}
\newcommand{\beas}{\begin{eqnarray*}}
\newcommand{\eeas}{\end{eqnarray*}}
\newcommand{\defi}{\stackrel{\rm def}{=}}
\newcommand{\non}{\nonumber}
\newcommand{\bquo}{\begin{quote}}
\newcommand{\enqu}{\end{quote}}
\renewcommand{\(}{\begin{equation}}
\renewcommand{\)}{\end{equation}}
\def \eqn#1#2{\begin{equation}#2\label{#1}\end{equation}}

\def\e{\epsilon}
\def\IZ{{\mathbb Z}}
\def\IR{{\mathbb R}}
\def\IC{{\mathbb C}}
\def\IQ{{\mathbb Q}}
\def\de{\partial}
\def\Tr{ \hbox{\rm Tr}}
\def\H{ \hbox{\rm H}}
\def\HE{ \hbox{$\rm H^{even}$}}
\def\HO{ \hbox{$\rm H^{odd}$}}
\def\K{ \hbox{\rm K}}
\def\Im{ \hbox{\rm Im}}
\def\Ker{ \hbox{\rm Ker}}
\def\const{\hbox {\rm const.}}
\def\o{\over}
\def\im{\hbox{\rm Im}}
\def\re{\hbox{\rm Re}}
\def\bra{\langle}\def\ket{\rangle}
\def\Arg{\hbox {\rm Arg}}
\def\Re{\hbox {\rm Re}}
\def\Im{\hbox {\rm Im}}
\def\exo{\hbox {\rm exp}}
\def\diag{\hbox{\rm diag}}
\def\longvert{{\rule[-2mm]{0.1mm}{7mm}}\,}
\def\a{\alpha}
\def\dag{{}^{\dagger}}
\def\tq{{\widetilde q}}
\def\p{{}^{\prime}}
\def\W{W}
\def\N{{\cal N}}
\def\hsp{,\hspace{.7cm}}

\def\br{\nonumber\\}
\def\IZ{{\mathbb Z}}
\def\IR{{\mathbb R}}
\def\IC{{\mathbb C}}
\def\IQ{{\mathbb Q}}
\def\IP{{\mathbb P}}
\def \eqn#1#2{\begin{equation}#2\label{#1}\end{equation}}

\newcommand{\C}{\ensuremath{\mathbb C}}
\newcommand{\Z}{\ensuremath{\mathbb Z}}
\newcommand{\R}{\ensuremath{\mathbb R}}
\newcommand{\rp}{\ensuremath{\mathbb {RP}}}
\newcommand{\cp}{\ensuremath{\mathbb {CP}}}
\newcommand{\vac}{\ensuremath{|0\rangle}}
\newcommand{\vact}{\ensuremath{|00\rangle}                    }
\newcommand{\oc}{\ensuremath{\overline{c}}}
\begin{titlepage}
\begin{flushright}
CHEP XXXXX
\end{flushright}
\bigskip
\def\thefootnote{\fnsymbol{footnote}}

\begin{center}
{\Large
{\bf $\epsilon$-Expansions Near Three Dimensions from \\  Conformal Field Theory \\ \vspace{0.1in} 
}
}
\end{center}

\bigskip
\begin{center}
{\large Pallab BASU$^a$\footnote{\texttt{pallabbasu@gmail.com}} and  Chethan KRISHNAN$^b$\footnote{\texttt{chethan.krishnan@gmail.com}} }
\vspace{0.1in}

\end{center}

\renewcommand{\thefootnote}{\arabic{footnote}}

\begin{center}
$^a$ {International Center for Theoretical Sciences,\\
IISc Campus, Bangalore 560012, India}\\
\vspace{0.2in}

$^b$ {Center for High Energy Physics,\\
Indian Institute of Science, Bangalore 560012, India}\\

\end{center}

\noindent
\begin{center} {\bf Abstract} \end{center}

We formally extend the CFT techniques introduced in arXiv: 1505.00963, to $\phi^{\frac{2d_0}{d_0-2}}$ theory in $d=d_0-\epsilon$ dimensions and use it to compute anomalous dimensions near $d_0=3, 4$ in a unified manner. We also do a similar analysis of the $O(N)$ model in three dimensions by developing a recursive combinatorial approach for OPE contractions.  
Our results match precisely with low loop perturbative computations. Finally, using 3-point correlators in the CFT, we comment on why the $\phi^3$ theory in $d_0=6$ is qualitatively different.


\vspace{1.6 cm}
\vfill

\end{titlepage}

\setcounter{footnote}{0}

\section{Introduction}

\noindent In a recent paper \cite{Rychkov} Rychkov and Tan have demonstrated that non-perturbative arguments can be used to determine the low loop anomalous dimensions of critical Wilson-Fisher theory in $d=4-\e$ dimensions. The argument is based purely on the idea that this theory is a conformal field theory, formalized via three (plus one\footnote{We count the assumption that the anomalous dimensions are analytic in the $\e \rightarrow 0$ limit, as a forth axiom.}) axioms. The fact that these results do not require perturbation theory is striking and worthy of further exploration. 

In this paper, we will apply the techniques of \cite{Rychkov} in $d=3-\e$ dimensions. In fact, we will begin with critical scalar field theory in $d=d_0-\epsilon$ dimensions and find that the approach allows a formal extension to general $d_0$ with $\phi^{\frac{2d_0}{d_0-2}}$ potential. In the end, because of various constraints, we will find that $d_0$ gets narrowed down to $4$ and $3$ -- $\phi^4$ in four dimensions and $\phi^6$ in three dimensions. For these cases the formalism allows a unified discussion. We will also find that $\phi^3$ in six dimensions does not allow a simple generalization of this idea. 

We further extend the analysis to the case of $O(N)$ model in three dimensions. One complication we have to deal with in $d_0=3$ $O(N)$ model is that the  OPE contractions required for the computations become too cumbersome. We therefore develop the recursive combinatorics of these contractions using a diagrammatic formalism. This approach might have some mileage even beyond the specific problem that we tackle here.

In all the cases, we find indeed that our results for the anomalous dimensions match precisely with extant results in the literature, where they overlap. As far as we are aware, the only analytical path to these results before this paper were via perturbative loop computations. The $\phi^6$ theories have been used to model multi-critical behavior, especially around tri-critical points.

Our results are based purely on constraints from three point functions. It seems plausible that these axioms, together with four-point functions and bootstrap equations might be constraining enough to determine the theory (more) completely \footnote{Some of the recent work on the conformal bootstrap is collected in \cite{Bootstrap}.
A pedagogical introduction can be found in \cite{R2}.}. We hope to come back to this question in the future.

\section{A Formal $\epsilon$-Expansion from Wilson-Fisher CFT}

We will consider scalar field theory in $d=d_0-\epsilon$ dimensions with the action
\bea
S=\int d^d x \left(\frac{1}{2} \partial \phi^2 + \frac{g}{\Gamma \big(1+\nu\big)} \mu ^{\alpha_0 \epsilon}\phi ^\nu \right)
\eea
where
\bea
\nu=\frac{2d_0}{d_0-2}, \ \ \alpha_0=\frac{2}{d_0-2}
\eea
One reason why this class of theories is interesting is because when $\epsilon \rightarrow 0$, ie., when $d=d_0$, the theory is renormalizable with a dimensionless coupling.  Another (related) reason, which is crucial from our perspective is that the theory has a weakly coupled fixed point at a coupling proportional to $\epsilon$, which we will call the Wilson-Fisher CFT \footnote{Typically, the case $d_0=4$ is called the Wilson-Fisher fixed point, but this is a natural generalization.}. When $\epsilon$ is finite, we have introduced the scale $\mu$ to make the coupling dimensionless. The action captures well-known $\phi^4$ theory in four dimensions (this was the case considered in \cite{Rychkov}), $\phi^6$ theory in three dimensions and $\phi^3$ theory in six dimensions. One goal of this paper is to present the discussion in a somewhat unified manner -- we will see that the CFT formalism goes through without hitch for the $d_0=3$ case as well. The $d_0=6$ $\epsilon$-expansion is known \cite{Macfarlane} to be  significantly different from the other two in its structure, the origins of this difference are immediate from the CFT perspective, as we will see. However, our CFT considerations based on 3-pt functions will only be able to make qualitative predictions about $d_0=6$. 

The dimensionality of the scalar in $d$-dimensions can be used to define the following quantities:
\bea
[\phi]\equiv \delta = \frac{d-2}{2}= \frac{d_0-2}{2}-\frac{\e}{2}
\eea
The Schwinger-Dyson equations of motion of the theory are given by
\bea
\Box \phi= \frac{g}{\Gamma \big(\nu\big)} \mu ^{\alpha_0 \epsilon}\phi ^{\nu-1}
\eea
Instead of viewing this as a dynamical equation, we will view this as a conformal multiplet {\em shortening condition} as in \cite{Rychkov}: in the free theory, $\phi ^{\nu-1}$ is a primary, but in the interacting theory it is defined by the LHS of the above equation, making it a descendant.  As in \cite{Rychkov} we will define our Wilson-Fisher theory by a set of three axioms. The first (Axiom I) of these says that the Wilson-Fisher theory is a conformal field theory. The second (Axiom II) says that operators $V_n$ and correlators between them in the Wilson-Fisher theory tend to operators $\phi^n$ and their correlators in the $\e \rightarrow 0$ (ie., free theory) limit. The third axiom is the most non-trivial one, and in our case it formalizes the multiplet shortening condition via the equality (Axiom III)
\bea
\Box V_1 = \alpha(\e) V_{\frac{d_0+2}{d_0-2}}, \label{axiom3}
\eea
where $\alpha(\e)$ is a-priori unknown. This means that the dimension of these operators are protected by the conformal algebra to be
\bea
\Delta_{\frac{d_0+2}{d_0-2}}=\Delta_1+2. \label{dim}
\eea
Note that in many of these statements, we need various integrality conditions on various functions of $d_0$ (like the subscript $\frac{d_0+2}{d_0-2}$ above) in order for them to make sense. The most stringent of them will turn out to be the condition that $2/(d_0-2)$ is a positive integer. Together with the condition that $d_0$ is an integer, it leaves only $d_0=3, 4$ as the solutions. We will discuss this when it arises, but we will proceed formally for now, for the simple reason that we can. 

The two-point function in the interacting CFT is
\bea
\langle V_1(x) V_1(y) \rangle = \frac{1}{|x-y|^{2 \Delta_1}} \label{WF2pt}
\eea
which in the free limit goes to
\bea
\langle \phi(x) \phi(y) \rangle = \frac{1}{|x-y|^{d_0-2}} \label{free2pt}
\eea
The scaling dimensions of $V_n$ is given by $\Delta_n= n\delta +\gamma_n$ where $\gamma_n$ is the anomalous dimension of $V_n$. Axiom II demands that the latter tend to the former in the free limit. We will assume further that the anomalous dimensions are analytic at $\e=0$ and admit a Taylor expansion\footnote{This is a major assumption, but it has the virtue that it seems to give the right answers as we will see. See also \cite{Rychkov}.} in $\e$:
\bea
\gamma_n=y_{n,1} \e + y_{n,2} \e^2+ ...
\eea
Now using
\bea
\Box_x\frac{1}{|x-y|^{2 \Delta_1}}=\frac{2 \Delta_1 (2\Delta_1+2 -d)}{|x-y|^{2 \Delta_1+2}},\hspace{0.7in} \label{once} \\
\Box_x\Box_y \frac{1}{|x-y|^{2 \Delta_1}}=\frac{4 \Delta_1 (\Delta_1+1) (2\Delta_1+2 -d)(2\Delta_1+4 -d)}{|x-y|^{2 \Delta_1+4}}, \label{twice}
\eea
and applying $\Box_x\Box_y$ on (\ref{WF2pt}), then using (\ref{axiom3}) and demanding that the result should tend to $\Box_x\Box_y$ acting on (\ref{free2pt}), we get the relation
\bea
\alpha(\e)=\sigma \sqrt{\frac{4d_0(d_0-2)\gamma_1}{\Gamma(\nu)}}
\eea
where we have extracted a sign $\sigma = \pm$ for the square root which will be fixed eventually via further CFT arguments. In arriving at the above result, we have used (\ref{axiom3}) and the fact that 
\bea
\langle V_{\frac{d_0+2}{d_0-2}}(x)V_{\frac{d_0+2}{d_0-2}}(y)\rangle \rightarrow \langle\phi^{\frac{d_0+2}{d_0-2}}(x)\phi^{\frac{d_0+2}{d_0-2}}(y)\rangle=\frac{\Gamma(\nu)}{|x-y|^{d_0+2}}.
\eea
The $\Gamma(\nu)$ arises because the full contraction of $\phi^k(x) \phi^k(y)$ gives rise to a $k!$, and for $k=\frac{d_0+2}{d_0-2}$, this can be written as $\Gamma(\nu)$. We follow \cite{Rychkov} closely in these steps. 

These further constraints arise from 3-pt correlators involving $V_n$ and $V_{n+1}$ \cite{Rychkov}. In the free theory limit, we can write
\bea
\phi^n(x) \times \phi^{n+1}(0) \supset f |x|^{-n(d_0-2)}\{ \phi(0) + \rho |x|^2 \phi^{\frac{d_0+2}{d_0-2}}(0)\}
\eea
which follows essentially from dimensional analysis. We will first determine the coefficients $f$ and $\rho$ that show up in this expression because we will need them.

\subsection{Counting Contractions}

The OPE coefficient $f$ can be trivially determined by direct contraction to be
\bea
f=(n+1)!
\eea
The coefficient $\rho$ requires a bit more work because it depends on $d_0$. To determine it, we first note that the number of contractions $(n-r)$ that one needs between $\phi^n$ and $\phi^{n+1}$, so that one is left with $\phi^{\frac{d_0+2}{d_0-2}}$ after the contractions, is given by 
\bea
(n + (n+1)) - 2 (n-r) = \frac{d_0+2}{d_0-2}.
\eea
This yields
\bea
r=\frac{2}{d_0-2}.
\eea
Now, of these $(n-r)$ contractions that need to be done, the first can be done by starting with $\phi$'s in $\phi^{n}$ and contracting with the $\phi$'s in $\phi^{n+1}$. A little thought shows that the choice of the $\phi$'s in $\phi^{n}$ can be made in ${}^nC_{n-r}$ ways, and the contractions with the $\phi^{n+1}$ can be done in $(n+1) \times n \times (n-1) \times ... \times (r+2)$ ways. So the net result for the number of contractions is
\bea
{}^nC_{n-r} \times \frac{(n+1)!}{(r+1)!}.
\eea
This quantity is equal to $\rho f$ because $f=(n+1)!$ is a common factor, so in the end we have
\bea
\rho(n)= \frac{{}^nC_{n-r}}{(r+1)!}.
\eea
In the case $d_0=4$ where $r=1$ this reduces to $n/2$ as was found in \cite{Rychkov}, and for the case $d_0=3$ where $r=2$, this yields
\bea
\rho_{d_0=3}=\frac{n(n-1)}{12}, \label{3rho}
\eea
which we will soon use to compute anomalous dimensions.

Note also the crucial role that the integrality of $r$ plays in these arguments. One might hope to generalize the conclusions to generic $r$ by re-writing the factorials in terms of Gamma functions, but the meaning of such an operation is unclear. This is because the arguments for the contractions were combinatorial. Indeed for $d_0=6$ were $r=1/2$, we will see that the situation is qualitatively different.

This is the first indication from the CFT approach that the $d_0=6$ case where $r$ is no longer integral is bound to have a conceptually different $\e$-expansion compared to the $d_0=3, 4$ cases. In particular, we will see that the latter theories have an anomalous dimension $\gamma_\phi$ that starts at $\mathcal{O} (\e^2)$ while the six dimensional theory it starts at $\mathcal{O} (\e)$.

\subsection{Matching with the Free Theory}

The idea now is to take 3-pt correlators involving the $V_n \times V_{n+1}$ OPEs and get constraints on the anomalous dimensions by demanding that they have a smooth free theory limit. The crucial point, as we emphasized in the discussion before (\ref{axiom3}), is that at finite $\epsilon$,  $V_{\frac{d_0+2}{d_0-2}}$ is no longer a primary. 

We are rather telegraphic in the discussion of this section (even though it is technically complete): we refer the reader to \cite{Rychkov} for more context and elaborations, this section is a direct generalization of their work.

The relevant terms in the OPE are \cite{Rychkov} (see also the original work of \cite{FGG1, FGG2, FGG3}):
\bea
V_n(x) \times V_{n+1} (x) \supset \tilde f |x|^{\Delta_1-\Delta_n-\Delta_{n+1}}(1+ q_1 x^\mu \partial_\mu+ q_2 x^\mu x^\nu \partial_\mu\partial_\nu +
q_3 x^2 \Box+...)V_1 (0) \label{2ptexpn}
\eea
We will demand that the (leading behavior of the) 3-pt correlators of this object tend to the corresponding free field 3-pt correlators 
\bea
\langle V_n(x) V_{n+1}(0) V_1(z)\rangle \rightarrow \langle \phi^n(x) \phi^{n+1}(0) \phi(z)\rangle \sim f |x|^{-n(d_0-2)}\langle\phi(0)\phi(z) \rangle, \hspace{0.6in} \label{first}\\
\langle V_n(x) V_{n+1}(0) V_{\frac{d_0+2}{d_0-2}}(z)\rangle \rightarrow \langle \phi^n(x) \phi^{n+1}(0) \phi^{\frac{d_0+2}{d_0-2}}(z)\rangle \sim f \rho |x|^{-n(d_0-2)+2}\langle\phi^{\frac{d_0+2}{d_0-2}}(0)\phi^{\frac{d_0+2}{d_0-2}}(z) \rangle, \nonumber \\ \label{second}
\eea
We are working here in the $|x| \ll |z|$ limit. The first line follows immediately from (\ref{2ptexpn}). To evaluate the LHS of the second line we use (\ref{2ptexpn}) and the fact that 
\bea
\langle V_1(0) V_{\frac{d_0+2}{d_0-2}}(z) \rangle = \alpha(\e)^{-1} \langle V_1(0) \Box V_{1}(z) \rangle = \frac{4 \alpha(\e)^{-1} \Delta_1 \gamma_1}{|z|^{2\Delta_1+2}} = \frac{4\sigma\Delta_1}{ |z|^{2\Delta_1+2}} \times \sqrt{\frac{\Gamma(\nu)\gamma_1}{4 d_0(d_0-2)}}.
\eea
where we have used (\ref{once}). The presence of $\sqrt{\gamma_1}$ suggests that this object vanishes in the $\e \rightarrow 0$ limit. Therefore, to reproduce (\ref{second}) we need $q_1$ and $q_2$ to stay finite in that limit. Noting that the box acting on the argument of $V_1(0)$ brings out a factor of $\alpha$ due to Axiom III (together with producing the requisite $V_{\frac{d_0+2}{d_0-2}}(z)$ inside the leftover 2-pt correlator), we find that for the correct free field match we need 
\bea
\lim_{\e \rightarrow 0} \ q_3 \alpha = \rho(n). \label{limit}
\eea
Using the expression (A.3) from \cite{Rychkov} for the $q_i$ we find that the $q_1, q_2$ finiteness conditions are automatically satisfied. Further, the leading behavior of $q_3$ in $\e \rightarrow 0$ limit comes from 
\bea
q_3 \approx \frac{\gamma_{n+1}-\gamma_{n}-\gamma_{1}}{4 d_0 \gamma_1} \label{q3def}
\eea
and so for $q_3$ to blow up, it is clearly a necessary condition that $y_{1,1}=0$,
\bea
\gamma_{1,1} \approx y_{1,2} \e^2.
\eea
This gives
\bea
\alpha\approx\e \ \sigma \sqrt{\frac{4d_0(d_0-2)y_{1,2}}{\Gamma(\nu)}}, \ \ 
q_3 \approx \frac{y_{n+1,1}-y_{n,1}}{4 d_0 y_{1,2} \epsilon}.
\eea
Putting them together we obtain the recursion relation 
\bea
y_{n+1,1}-y_{n,1}=\sigma\sqrt{ \frac{4d_0\Gamma(\nu) y_{1,2}}{d_0-2}}\ \rho(n) \equiv K \rho(n)
\eea
Summing the telescoping series, we get 
\bea
y_{n,1}=K\sum_{m=1}^{n-1} \rho(m) \label{ans}
\eea
which is the final answer, once we fix the numerical value of $K$ (which is the same as fixing the numerical value of $y_{1,2}$). This can be accomplished via (\ref{dim}), which can be written as
\bea
\frac{d_0+2}{d_0-2} \left(\frac{d}{2}-1\right)+\gamma_{\frac{d_0+2}{d_0-2}}=\left(\frac{d}{2}-1\right)+\gamma_1 +2.
\eea
To leading power in $\e$, this translates to
\bea
y_{\frac{d_0+2}{d_0-2},1}=\frac{2}{d_0-2}. 
\eea
Now this can be used to fix $K$ by setting $n=\frac{d_0+2}{d_0-2}$ in (\ref{ans}). For $d_0=4$, this gives $K=2/3$ and using this one fixes $\sigma=+$ and $y_{1,2}=1/108$, reproducing the results of \cite{Rychkov}\footnote{Note that our definition of $K$ is slightly different from theirs.}. For $d_0=3$, using (\ref{3rho}) we get $K=6/5$. The final answers written directly in terms of anomalous dimensions are
\bea
\gamma^{d_0=3}_\phi= \frac{\e^2}{1000} +\mathcal{O}(\e^3), \ \  \gamma^{d_0=3}_{\phi^n}=\frac{1}{30}n(n-1)(n-2) \e +\mathcal{O}(\e^2).
\eea
The formulas for $\sum_{m=1}^n m$ and $\sum_{m=1}^n m^2$ are useful in getting these results. 
The final result agrees with the perturbative results in (for example) \cite{Hager} where they overlap. 

\subsection{Fixing Loose Ends} 

In obtaining the above result, we summed the telescoping series, and for doing that we implicitly assumed that the recursion relations arising from the OPEs involving  the descendants has the same form as the ones arising from primaries. This needs an explicit check for $n=4, 5$, because these are the only cases where the contractions involve descendants as well. This check can be done using relations (A.4-A.7) in \cite{Rychkov}. 

Another assumption we made is that $y_{1,2}\neq 0$. To prove this, we first note that the $q_3 \sim 1/\sqrt{\gamma_1}$ due to (\ref{limit}). Using (\ref{q3def}) for $n=1,2,3,4$, this gives
\bea
\frac{1}{\sqrt{\gamma_1}} \sim \frac{\gamma_2}{\gamma_1}, \ \ \frac{1}{\sqrt{\gamma_1}} \sim \frac{\gamma_3-\gamma_2}{\gamma_1}, \ \ \frac{1}{\sqrt{\gamma_1}} \sim \frac{\gamma_4-\gamma_3}{\gamma_1}, \ \ \frac{1}{\sqrt{\gamma_1}} \sim \frac{2\e-\gamma_4}{\gamma_1} 
\eea  
It is straightforward to check that these relations can all hold together at the same time, only if  $\gamma_1 \sim \e^2$. (Note that when one adds the last three conditions above, the resultant relation together with the first, gives rise to a system that is identical to that discussed near eqn.(3.39) in \cite{Rychkov}.)

The arguments in this subsection apply without any further subtleties to the $O(N)$ model that we discuss in the next section, so we will not repeat this discussion there.


\section{Generalization to $O(N)$ Model}

Now we will consider generalization of the previous discussion to the $O(N)$ model in $d_0=3$. The Lagrangian is of the form
\bea
L=\int d^{3-\e}x \left(\frac{1}{2} \partial \vec{\phi}^2 + \frac{g}{6!} \mu ^{2 \epsilon}(\vec{\phi}^2)^3 \right)
\eea
where $\vec{\phi} \equiv \phi^a$ stands for a collection of $N$ scalar fields indexed by $a$. The theory has an $O(N)$ symmetry. We will use the techniques of \cite{Rychkov} to compute the anomalous dimensions of two series of operators in this CFT
\bea
W^a_{2p+1} \ {\rm and}  \ W_{2p}
\eea
which tend to the free field operators
\bea
\Phi^a_{2p+1}\equiv \phi^a(\vec{\phi}^2)^p,  \ {\rm and} \ \Phi_{2p}\equiv (\vec{\phi}^2)^p
\eea
in the $\e \rightarrow 0$ limit. Apart from the relation
\bea
\Box W^a_1= \alpha W^a_5
\eea
which makes $W^a_5$ a descendant, the $W$ operators are all primaries. Evaluating the left and right sides of 
\bea
\langle \Box_x W_1^a(x) \Box_y W_1^b(y) \rangle=\alpha^2 \langle W_5^a(x) W_5^b(y \rangle
\eea
independently in the free limit, parallel to the discussion in the previous section, we find that 
\bea
\alpha=\sigma \sqrt{\frac{3 \gamma_1}{2(2+N)(4+N)}}, \label{ONalpha}
\eea
where $\sigma$ is again a sign that will soon be determined. The $N$-dependence arises from the various ways that $\Phi^a_{5}(x)$ can be contracted with $\Phi^a_{5}(y)$ in the free theory. This is the first in a series of contractions that we will need -- in this particular case it can be done by inspection.

We will fix the anomalous dimensions by constructing telescoping series as in the last section. The relevant relations that can be used to determine these series are
\bea
\Phi_{2p}(x) \times \Phi^a_{2p+1}(0) \supset f_{2p} |x|^{-2p}\{ \Phi^a_1(0)+\rho_{2p} |x|^2 \Phi^a_{5}(0) \}, \\
\Phi^a_{2p+1}(x) \times \Phi_{2p+2}(0) \supset f_{2p+1} |x|^{-2p-1}\{ \Phi^a_1(0)+\rho_{2p+1} |x|^2 \Phi^a_{5}(0) \}
\eea
To determine the coefficients $\rho$ which are crucial for proceeding further, it behooves us to develop a formalism which can accomplish contractions systematically. This formalism might be of some use/interest in and of itself, so this is what we turn to next.

\subsection{Counting Contractions Using Cow-Pies}

We will develop a recursive approach to compute the coefficients $f$ and $\rho$. To do this we first introduce some graphical notation. We first define
\bea
F^{p,r}_{p+q,s;m}
\eea
to stand for the total number of contractions between $(\phi^2)^p \phi^{\mu_1} ...\phi^{\mu_r}$ and   $(\phi^2)^{p+q} \phi^{\mu_1} ...\phi^{\mu_s}$ such that $m$ of the $\phi$'s are left uncontracted.

Graphically, we describe this using a cow-pie diagram, as shown in the next figure. $F^{p,r}_{p+q,s;m}$, in this language, stands for the total number of ways in which the kernels in the upper array of cow-pies in the figure
\begin{figure}[h]
\begin{tikzpicture}

\draw (2,-0.25) node[anchor=south] {\textbullet\quad\textbullet\quad\textbullet};
\draw (7,-0.25) node[anchor=south] {\textbullet\textbullet\textbullet};
\begin{scope}
    \draw (0,0) ellipse (1cm and 0.5cm);
    \node[draw=red] at (-0.4,0)  {$\times$};
    \node[draw=red] at (0.4,0)  {$\times$} ;
\end{scope}

\begin{scope}[xshift=4cm]
\draw (0,0) ellipse (1cm and 0.5cm);
\node[draw=red] at (-0.4,0)  {$\times$};
\node[draw=red] at (0.4,0)  {$\times$} ;
\end{scope}

\begin{scope}[xshift=6cm]
\draw (0,0) circle (0.5cm);
\node[draw=red] at (0,0)  {$\times$} ;
\end{scope}

\begin{scope}[xshift=8cm]
\draw (0,0) circle (0.5cm);
\node[draw=red] at (0,0)  {$\times$} ;
\end{scope}

\draw (2,-2.25) node[anchor=south] {\textbullet\quad\textbullet\quad\textbullet};
\draw (6,-2.25) node[anchor=south] {\textbullet\quad\textbullet\quad\textbullet};
\draw (11,-2.25) node[anchor=south] {\textbullet\textbullet\textbullet};

\begin{scope}[yshift=-2cm]
\draw (0,0) ellipse (1cm and 0.5cm);
\node[draw=red] at (-0.4,0)  {$\times$};
\node[draw=red] at (0.4,0)  {$\times$} ;
\end{scope}

\begin{scope}[xshift=4cm,yshift=-2cm]
\draw (0,0) ellipse (1cm and 0.5cm);
\node[draw=red] at (-0.4,0)  {$\times$};
\node[draw=red] at (0.4,0)  {$\times$} ;
\end{scope}
 
\begin{scope}[xshift=8cm,yshift=-2cm]
\draw (0,0) ellipse (1cm and 0.5cm);
\node[draw=red] at (-0.4,0)  {$\times$};
\node[draw=red] at (0.4,0)  {$\times$} ;
\end{scope}

\begin{scope}[xshift=10cm,yshift=-2cm]
\draw (0,0) circle (0.5cm);
\node[draw=red] at (0,0)  {$\times$} ;
\end{scope}

\begin{scope}[xshift=12cm,yshift=-2cm]
\draw (0,0) circle (0.5cm);
\node[draw=red] at (0,0)  {$\times$} ;
\end{scope}

\begin{scope}[yshift=0.7cm]
\draw [thick,decorate,decoration={brace,amplitude=10pt}] (-1.0,0) -- (5.0,0);
\node[] at (2.05,0.55) {\large $p$}; 
\end{scope}

\begin{scope}[yshift=0.7cm]
\draw [thick,decorate,decoration={brace,amplitude=10pt}] (5.5,0) -- (8.5,0);
\node[] at (7.05,0.55) {\large $r$}; 
\end{scope}

\begin{scope}[yshift=-2.7cm]
\draw [thick,decorate,decoration={brace,mirror,amplitude=10pt}] (-1.0,0) -- (9.0,0);
\node[] at (4.0,-.6) {\large $p+q$}; 
\end{scope}

\begin{scope}[yshift=-2.7cm]
\draw [thick,decorate,decoration={brace,mirror,amplitude=10pt}] (9.5,0) -- (12.5,0);
\node[] at (11.0,-.6) {\large $s$}; 
\end{scope}

\end{tikzpicture}
\end{figure}
can be contracted (aka connected by line-segments) with the kernels in the lower array of cow-pies -- but with the restriction that one has a total of $m$ leftover un-contracted kernels. We use the following terminology in what follows -- in the figure, the upper array contains $p$ double cow-pies and $r$ single cow-pies, while the lower array contains $p+q$  double cow-pies and $s$ single cow-pies. 

The rationale behind the introduction of this notation is that the quantities we want to compute can be seen  to be
\bea
f_{2p}=F^{p,0}_{p,1;1}, \ \ f_{2p+1}=F^{p,1}_{p+1,0;1}, \ \ f_{2p}\rho_{2p}=F^{p,0}_{p,1;5}, \ \ f_{2p+1}\rho_{2p+1}=F^{p,1}_{p+1,0;5}.
\eea
We will evaluate these quantities by setting up a descending iteration in $p$. 

We will start by evaluating  $F^{p,0}_{p,1;1}$. There are three distinct kinds of contractions one encounters when starting from $F^{p,0}_{p,1;1}$ and trying to reduce $p$ recursively. The idea is that we try to count the number of ways in which the $p$'th upper double cow-pie  (PUDC, for short) can be contracted with the lower array. These can be symbolized by the following three figures:
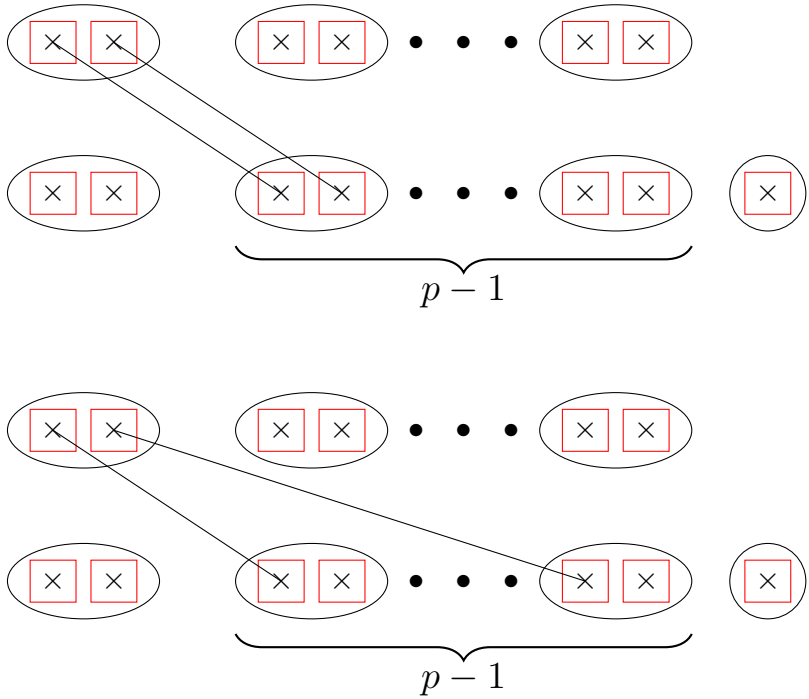
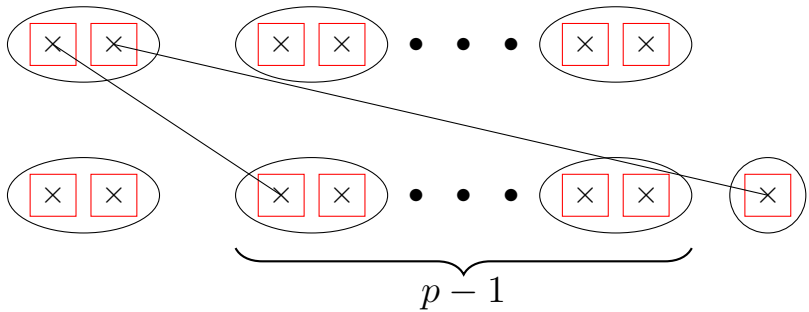
\begin{figure}[h]
\begin{tikzpicture}

\draw (5,-0.25) node[anchor=south] {\textbullet\quad\textbullet\quad\textbullet};

\begin{scope}
    \draw (0,0) ellipse (1cm and 0.5cm);
    \node[draw=red] at (-0.4,0)  {$\times$};
    \node[draw=red] at (0.4,0)  {$\times$} ;
\end{scope}

\begin{scope}[xshift=3cm]
\draw (0,0) ellipse (1cm and 0.5cm);
\node[draw=red] at (-0.4,0)  {$\times$};
\node[draw=red] at (0.4,0)  {$\times$} ;
\end{scope}
 
\begin{scope}[xshift=7cm]
\draw (0,0) ellipse (1cm and 0.5cm);
\node[draw=red] at (-0.4,0)  {$\times$};
\node[draw=red] at (0.4,0)  {$\times$} ;
\end{scope}

\draw (5,-2.25) node[anchor=south] {\textbullet\quad\textbullet\quad\textbullet};

\begin{scope}[yshift=-2cm]
\draw (0,0) ellipse (1cm and 0.5cm);
\node[draw=red] at (-0.4,0)  {$\times$};
\node[draw=red] at (0.4,0)  {$\times$} ;
\end{scope}

\begin{scope}[xshift=3cm,yshift=-2cm]
\draw (0,0) ellipse (1cm and 0.5cm);
\node[draw=red] at (-0.4,0)  {$\times$};
\node[draw=red] at (0.4,0)  {$\times$} ;
\end{scope}
 
\begin{scope}[xshift=7cm,yshift=-2cm]
\draw (0,0) ellipse (1cm and 0.5cm);
\node[draw=red] at (-0.4,0)  {$\times$};
\node[draw=red] at (0.4,0)  {$\times$} ;
\end{scope}

\begin{scope}[xshift=9cm,yshift=-2cm]
\draw (0,0) circle (0.5cm);
\node[draw=red] at (0,0)  {$\times$} ;

\end{scope}

\draw (-0.4,0) -- (2.6,-2);
\draw (0.4,0) --  (3.4,-2);

\begin{scope}[yshift=-2.7cm]
\draw [thick,decorate,decoration={brace,mirror,amplitude=10pt}] (2,0) -- (8.0,0);
\node[] at (5.0,-.6) {\large $p-1$}; 
\end{scope}

\end{tikzpicture}
\end{figure}

\begin{figure}[h]
\begin{tikzpicture}

\draw (5,-0.25) node[anchor=south] {\textbullet\quad\textbullet\quad\textbullet};

\begin{scope}
    \draw (0,0) ellipse (1cm and 0.5cm);
    \node[draw=red] at (-0.4,0)  {$\times$};
    \node[draw=red] at (0.4,0)  {$\times$} ;
\end{scope}

\begin{scope}[xshift=3cm]
\draw (0,0) ellipse (1cm and 0.5cm);
\node[draw=red] at (-0.4,0)  {$\times$};
\node[draw=red] at (0.4,0)  {$\times$} ;
\end{scope}
 
\begin{scope}[xshift=7cm]
\draw (0,0) ellipse (1cm and 0.5cm);
\node[draw=red] at (-0.4,0)  {$\times$};
\node[draw=red] at (0.4,0)  {$\times$} ;
\end{scope}

\draw (5,-2.25) node[anchor=south] {\textbullet\quad\textbullet\quad\textbullet};

\begin{scope}[yshift=-2cm]
\draw (0,0) ellipse (1cm and 0.5cm);
\node[draw=red] at (-0.4,0)  {$\times$};
\node[draw=red] at (0.4,0)  {$\times$} ;
\end{scope}

\begin{scope}[xshift=3cm,yshift=-2cm]
\draw (0,0) ellipse (1cm and 0.5cm);
\node[draw=red] at (-0.4,0)  {$\times$};
\node[draw=red] at (0.4,0)  {$\times$} ;
\end{scope}
 
\begin{scope}[xshift=7cm,yshift=-2cm]
\draw (0,0) ellipse (1cm and 0.5cm);
\node[draw=red] at (-0.4,0)  {$\times$};
\node[draw=red] at (0.4,0)  {$\times$} ;
\end{scope}

\begin{scope}[xshift=9cm,yshift=-2cm]
\draw (0,0) circle (0.5cm);
\node[draw=red] at (0,0)  {$\times$} ;

\end{scope}

\draw (-0.4,0) -- (2.6,-2);
\draw (0.4,0) --  (6.6,-2);

\begin{scope}[yshift=-2.7cm]
\draw [thick,decorate,decoration={brace,mirror,amplitude=10pt}] (2,0) -- (8.0,0);
\node[] at (5.0,-.6) {\large $p-1$}; 
\end{scope}

\end{tikzpicture}
\end{figure}

\begin{figure}[h]
\begin{tikzpicture}

\draw (5,-0.25) node[anchor=south] {\textbullet\quad\textbullet\quad\textbullet};

\begin{scope}
    \draw (0,0) ellipse (1cm and 0.5cm);
    \node[draw=red] at (-0.4,0)  {$\times$};
    \node[draw=red] at (0.4,0)  {$\times$} ;
\end{scope}

\begin{scope}[xshift=3cm]
\draw (0,0) ellipse (1cm and 0.5cm);
\node[draw=red] at (-0.4,0)  {$\times$};
\node[draw=red] at (0.4,0)  {$\times$} ;
\end{scope}
 
\begin{scope}[xshift=7cm]
\draw (0,0) ellipse (1cm and 0.5cm);
\node[draw=red] at (-0.4,0)  {$\times$};
\node[draw=red] at (0.4,0)  {$\times$} ;
\end{scope}

\draw (5,-2.25) node[anchor=south] {\textbullet\quad\textbullet\quad\textbullet};

\begin{scope}[yshift=-2cm]
\draw (0,0) ellipse (1cm and 0.5cm);
\node[draw=red] at (-0.4,0)  {$\times$};
\node[draw=red] at (0.4,0)  {$\times$} ;
\end{scope}

\begin{scope}[xshift=3cm,yshift=-2cm]
\draw (0,0) ellipse (1cm and 0.5cm);
\node[draw=red] at (-0.4,0)  {$\times$};
\node[draw=red] at (0.4,0)  {$\times$} ;
\end{scope}
 
\begin{scope}[xshift=7cm,yshift=-2cm]
\draw (0,0) ellipse (1cm and 0.5cm);
\node[draw=red] at (-0.4,0)  {$\times$};
\node[draw=red] at (0.4,0)  {$\times$} ;
\end{scope}

\begin{scope}[xshift=9cm,yshift=-2cm]
\draw (0,0) circle (0.5cm);
\node[draw=red] at (0,0)  {$\times$} ;

\end{scope}

\draw (-0.4,0) -- (2.6,-2);
\draw (0.4,0) --  (9.0,-2);

\begin{scope}[yshift=-2.7cm]
\draw [thick,decorate,decoration={brace,mirror,amplitude=10pt}] (2,0) -- (8.0,0);
\node[] at (5.0,-.6) {\large $p-1$}; 
\end{scope}

\end{tikzpicture}
\end{figure}

It is easy to see that there are $2 \times p \times N$ ways of contracting the PUDC the first way\footnote{The $N$ arises because a closed loop of contractions is a trace of the form $\delta^{aa}$ in terms of the $O(N)$ indices.}, while there are $2p \times 2 (p-1)$ ways of doing the second type of contractions, and there are $2p \times 2$ ways of doing the contractions the third way. Note that in each case, a bit of thought reveals that the result of each type of contraction is simply $F^{p-1,0}_{p-1,1,1}$. So we get a recursion relation 
\bea
F^{p,0}_{p,1;1}=(2 p N+4p(p-1)+4p) F^{p-1,0}_{p-1,1;1} \equiv (2 p +N)\times 2 p \times F^{p-1,0}_{p-1,1;1}
\eea
Together with the knowledge that $F^{0,0}_{0,1;1}=1$ (which follows trivially upon inspection) this immediately lets us evaluate
\bea
f_{2p}\equiv F^{p,0}_{p,1;1}=(2p+N) \times (2p) \times ... \times (2+N) \times 2.
\eea
An entirely similar recursion can be constructed for $F^{p,1}_{p+1,0;1}$, T
and a closely related result follows: 
\bea
F^{p,1}_{p+1,0;1}=2(p+1) \times (2 p + N) \times F^{p-1,0}_{p-1,1;1}
\eea
The launching condition for the iteration is seen by inspection to be $F^{0,1}_{1,0;1}=2$. This yields
\bea
f_{2p+1}\equiv F^{p,1}_{p+1,0;1}=(2p+2) \times (2p+N) \times ... \times 4 \times (2+N) \times 2.
\eea 

The results for $f$'s are sufficiently simple that it is possible to guess these answers by doing the contractions explicitly (if somewhat painfully) for low $p$'s. So our recursive formalism might seem like an overkill. However, the usefulness of the formalism  becomes clear in evaluating the $\rho$'s (or equivalently $F^{p,0}_{p,1;5}$ and $F^{p,1}_{p+1,0;5}$) for which we have not been able to come up with an alternate way to count the contractions without using the recursion relations\footnote{In hindsight, it seems plausible that one can perhaps guess the right expressions for $\rho$ by matching with the $N=1$ case, as well as some general arguments about the order of polynomials that one can expect (in $p$ and $N$) and explicitly working out the low order cases to match undetermined coefficients. This is ugly and feels like cheating, so we will stick to our systematic combinatorial approach, which has its own elegance. This enables us to use the match with the $N=1$ case as a sanity check on our results.}

We will start with $F^{p,0}_{p,1;5}$. There are three distinct types of contractions one needs to take care of in this case. The first corresponds to the case where both kernels in the PUDC are contracted (Type I), the second corresponds to only one of the PUDC kernels being contracted (Type II), and the third corresponds to none of the PUDC kernels being contracted (Type III). Type I follows a very similar structure as the previous cases we considered and contributes $2 p \times (2p+N) \times  F^{p-1,0}_{p-1,1;5}$ to the right hand side of the iteration equation, we will skip the details and the associated figures. Type II on the other hand splits into two subcases which can be captured by the following figures:
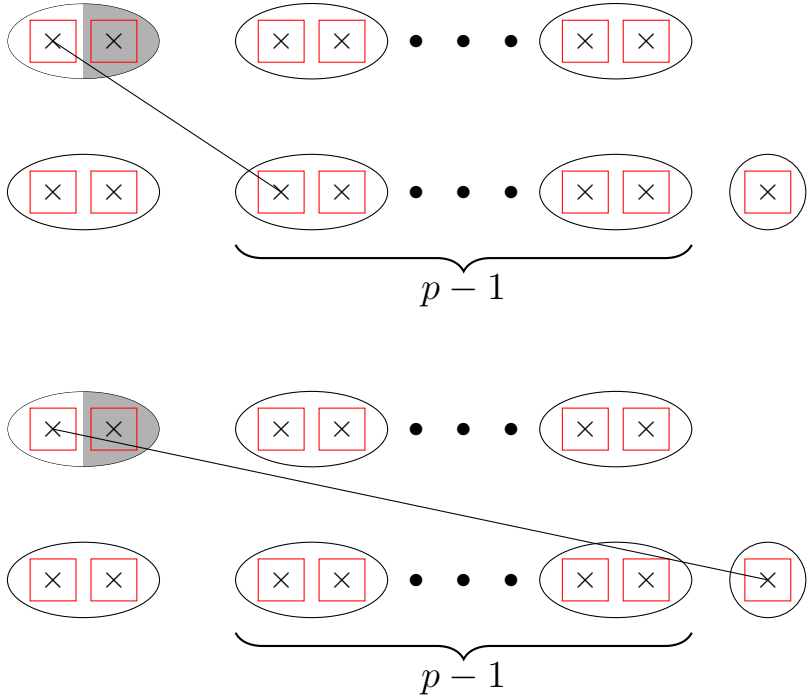
\begin{figure}[h]
\begin{tikzpicture}

\draw (5,-0.25) node[anchor=south] {\textbullet\quad\textbullet\quad\textbullet};

\begin{scope}
    \clip (0,0) ellipse (1cm and 0.5cm);
    \fill[black!30] (0cm,-1cm) rectangle (1cm, 1cm);
    \draw (0,0) ellipse (1cm and 0.5cm);
    \node[draw=red] at (-0.4,0)  {$\times$};
    \node[draw=red] at (0.4,0)  {$\times$} ;
\end{scope}

\begin{scope}[xshift=3cm]
\draw (0,0) ellipse (1cm and 0.5cm);
\node[draw=red] at (-0.4,0)  {$\times$};
\node[draw=red] at (0.4,0)  {$\times$} ;
\end{scope}
 
\begin{scope}[xshift=7cm]
\draw (0,0) ellipse (1cm and 0.5cm);
\node[draw=red] at (-0.4,0)  {$\times$};
\node[draw=red] at (0.4,0)  {$\times$} ;
\end{scope}

\draw (5,-2.25) node[anchor=south] {\textbullet\quad\textbullet\quad\textbullet};

\begin{scope}[yshift=-2cm]
\draw (0,0) ellipse (1cm and 0.5cm);
\node[draw=red] at (-0.4,0)  {$\times$};
\node[draw=red] at (0.4,0)  {$\times$} ;
\end{scope}

\begin{scope}[xshift=3cm,yshift=-2cm]
\draw (0,0) ellipse (1cm and 0.5cm);
\node[draw=red] at (-0.4,0)  {$\times$};
\node[draw=red] at (0.4,0)  {$\times$} ;
\end{scope}
 
\begin{scope}[xshift=7cm,yshift=-2cm]
\draw (0,0) ellipse (1cm and 0.5cm);
\node[draw=red] at (-0.4,0)  {$\times$};
\node[draw=red] at (0.4,0)  {$\times$} ;
\end{scope}

\begin{scope}[xshift=9cm,yshift=-2cm]
\draw (0,0) circle (0.5cm);
\node[draw=red] at (0,0)  {$\times$} ;

\end{scope}

\draw (-0.4,0) -- (2.6,-2);

\begin{scope}[yshift=-2.7cm]
\draw [thick,decorate,decoration={brace,mirror,amplitude=10pt}] (2,0) -- (8.0,0);
\node[] at (5.0,-.6) {\large $p-1$}; 
\end{scope}

\end{tikzpicture}
\end{figure}

\begin{figure}[h]
\begin{tikzpicture}

\draw (5,-0.25) node[anchor=south] {\textbullet\quad\textbullet\quad\textbullet};

\begin{scope}
    \clip (0,0) ellipse (1cm and 0.5cm);
    \fill[black!30] (0cm,-1cm) rectangle (1cm, 1cm);
    \draw (0,0) ellipse (1cm and 0.5cm);
    \node[draw=red] at (-0.4,0)  {$\times$};
    \node[draw=red] at (0.4,0)  {$\times$} ;
\end{scope}

\begin{scope}[xshift=3cm]
\draw (0,0) ellipse (1cm and 0.5cm);
\node[draw=red] at (-0.4,0)  {$\times$};
\node[draw=red] at (0.4,0)  {$\times$} ;
\end{scope}
 
\begin{scope}[xshift=7cm]
\draw (0,0) ellipse (1cm and 0.5cm);
\node[draw=red] at (-0.4,0)  {$\times$};
\node[draw=red] at (0.4,0)  {$\times$} ;
\end{scope}

\draw (5,-2.25) node[anchor=south] {\textbullet\quad\textbullet\quad\textbullet};

\begin{scope}[yshift=-2cm]
\draw (0,0) ellipse (1cm and 0.5cm);
\node[draw=red] at (-0.4,0)  {$\times$};
\node[draw=red] at (0.4,0)  {$\times$} ;
\end{scope}

\begin{scope}[xshift=3cm,yshift=-2cm]
\draw (0,0) ellipse (1cm and 0.5cm);
\node[draw=red] at (-0.4,0)  {$\times$};
\node[draw=red] at (0.4,0)  {$\times$} ;
\end{scope}
 
\begin{scope}[xshift=7cm,yshift=-2cm]
\draw (0,0) ellipse (1cm and 0.5cm);
\node[draw=red] at (-0.4,0)  {$\times$};
\node[draw=red] at (0.4,0)  {$\times$} ;
\end{scope}

\begin{scope}[xshift=9cm,yshift=-2cm]
\draw (0,0) circle (0.5cm);
\node[draw=red] at (0,0)  {$\times$} ;

\end{scope}

\draw (-0.4,0) -- (9,-2);

\begin{scope}[yshift=-2.7cm]
\draw [thick,decorate,decoration={brace,mirror,amplitude=10pt}] (2,0) -- (8.0,0);
\node[] at (5.0,-.6) {\large $p-1$}; 
\end{scope}

\end{tikzpicture}
\end{figure}
The shaded kernel emphasizes the fact that it must remain un-contracted and that the rest of the contractions are only among the remaining kernels. 
A bit of thought shows that the first of these figures can be seen to be equal to $4p \times F^{p-1,0}_{p-1,2;4}$, and that the second one is equal to $2 \times F^{p-1,0}_{p,0;4}$, so together Type II makes a contribution of$4p \times F^{p-1,0}_{p-1,2;4}+2 \times F^{p-1,0}_{p,0;4}$ to the right hand side of the iteration relation for $F^{p,0}_{p,1;5}$. 

\newpage
Turning to Type III, the figure takes the form
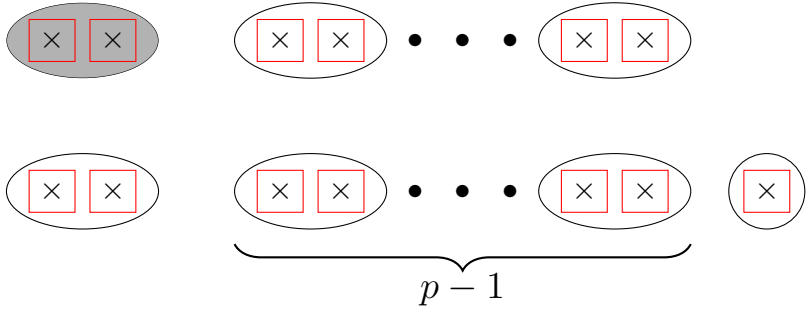
\begin{figure}[h]
\begin{tikzpicture}

\draw (5,-0.25) node[anchor=south] {\textbullet\quad\textbullet\quad\textbullet};

\begin{scope}
    \clip (0,0) ellipse (1cm and 0.5cm);
    \fill[black!30] (-1cm,-1cm) rectangle (1cm, 1cm);
    \draw (0,0) ellipse (1cm and 0.5cm);
    \node[draw=red] at (-0.4,0)  {$\times$};
    \node[draw=red] at (0.4,0)  {$\times$} ;
\end{scope}

\begin{scope}[xshift=3cm]
\draw (0,0) ellipse (1cm and 0.5cm);
\node[draw=red] at (-0.4,0)  {$\times$};
\node[draw=red] at (0.4,0)  {$\times$} ;
\end{scope}
 
\begin{scope}[xshift=7cm]
\draw (0,0) ellipse (1cm and 0.5cm);
\node[draw=red] at (-0.4,0)  {$\times$};
\node[draw=red] at (0.4,0)  {$\times$} ;
\end{scope}

\draw (5,-2.25) node[anchor=south] {\textbullet\quad\textbullet\quad\textbullet};

\begin{scope}[yshift=-2cm]
\draw (0,0) ellipse (1cm and 0.5cm);
\node[draw=red] at (-0.4,0)  {$\times$};
\node[draw=red] at (0.4,0)  {$\times$} ;
\end{scope}

\begin{scope}[xshift=3cm,yshift=-2cm]
\draw (0,0) ellipse (1cm and 0.5cm);
\node[draw=red] at (-0.4,0)  {$\times$};
\node[draw=red] at (0.4,0)  {$\times$} ;
\end{scope}
 
\begin{scope}[xshift=7cm,yshift=-2cm]
\draw (0,0) ellipse (1cm and 0.5cm);
\node[draw=red] at (-0.4,0)  {$\times$};
\node[draw=red] at (0.4,0)  {$\times$} ;
\end{scope}

\begin{scope}[xshift=9cm,yshift=-2cm]
\draw (0,0) circle (0.5cm);
\node[draw=red] at (0,0)  {$\times$} ;

\end{scope}


\begin{scope}[yshift=-2.7cm]
\draw [thick,decorate,decoration={brace,mirror,amplitude=10pt}] (2,0) -- (8.0,0);
\node[] at (5.0,-.6) {\large $p-1$}; 
\end{scope}

\end{tikzpicture}
\end{figure}

This is simply a contribution of $F^{p-1,0}_{p,1;3}$ to the right hand side of the iteration relation for $F^{p,0}_{p,1;5}$. Altogether then, the iteration relation for $F^{p,0}_{p,1;5}$ takes the form
\bea
F^{p,0}_{p,1;5}=2 p (2p+N) \times  F^{p-1,0}_{p-1,1;5}+4p \times F^{p-1,0}_{p-1,2;4}+2 \times F^{p-1,0}_{p,0;4}+F^{p-1,0}_{p,1;3}. \label{even1}
\eea
Unlike in the previous case of $f$'s we see that now there are new structures arising on the right hand side. So we need to come up with recursion relations for them as well. When we have a closed system of recursion relations, we will have enough information to solve for all of them. So now we turn to the recursion relations for $F^{p,0}_{p,2;4}, F^{p,0}_{p+1,0;4}$ and $F^{p,0}_{p+1,1;3}$. 

For $F^{p,0}_{p,2;4}$ there are two types of contractions for the PUDC with the lower layer cow-pies. Type I, which has both kernels of PUDC contracted, and Type II which has only one kernel of PUDC contracted. There is no Type III because it is easy to convince oneself that when both kernels of PUDC are un-contracted, the result must give zero. 

Type I gets contributions from four types of figures. Of these the first two are familiar structures that we have seen before leading to the contribution $2 N p + 4p(p-1)) \times F^{p-1,0}_{p-1,2;4}$, and the third one also works along similar lines adding a contribution $8p \times F^{p-1,0}_{p-1,2;4}$. The forth figure takes the form: 
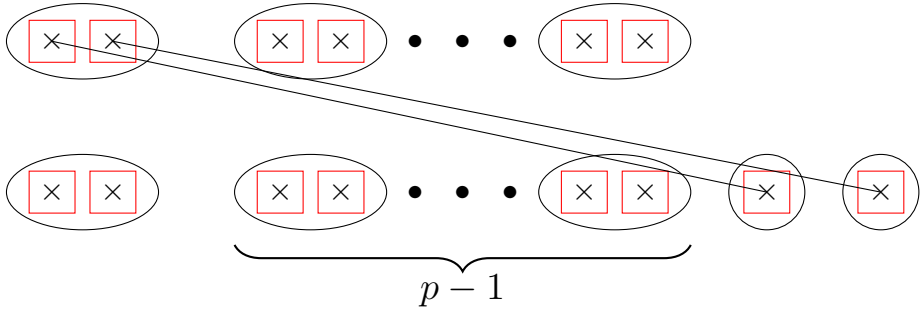
\begin{figure}[h]
\begin{tikzpicture}

\draw (5,-0.25) node[anchor=south] {\textbullet\quad\textbullet\quad\textbullet};

\begin{scope}
    \draw (0,0) ellipse (1cm and 0.5cm);
    \node[draw=red] at (-0.4,0)  {$\times$};
    \node[draw=red] at (0.4,0)  {$\times$} ;
\end{scope}

\begin{scope}[xshift=3cm]
\draw (0,0) ellipse (1cm and 0.5cm);
\node[draw=red] at (-0.4,0)  {$\times$};
\node[draw=red] at (0.4,0)  {$\times$} ;
\end{scope}
 
\begin{scope}[xshift=7cm]
\draw (0,0) ellipse (1cm and 0.5cm);
\node[draw=red] at (-0.4,0)  {$\times$};
\node[draw=red] at (0.4,0)  {$\times$} ;
\end{scope}

\draw (5,-2.25) node[anchor=south] {\textbullet\quad\textbullet\quad\textbullet};

\begin{scope}[yshift=-2cm]
\draw (0,0) ellipse (1cm and 0.5cm);
\node[draw=red] at (-0.4,0)  {$\times$};
\node[draw=red] at (0.4,0)  {$\times$} ;
\end{scope}

\begin{scope}[xshift=3cm,yshift=-2cm]
\draw (0,0) ellipse (1cm and 0.5cm);
\node[draw=red] at (-0.4,0)  {$\times$};
\node[draw=red] at (0.4,0)  {$\times$} ;
\end{scope}
 
\begin{scope}[xshift=7cm,yshift=-2cm]
\draw (0,0) ellipse (1cm and 0.5cm);
\node[draw=red] at (-0.4,0)  {$\times$};
\node[draw=red] at (0.4,0)  {$\times$} ;
\end{scope}

\begin{scope}[xshift=9cm,yshift=-2cm]
\draw (0,0) circle (0.5cm);
\node[draw=red] at (0,0)  {$\times$} ;

\end{scope}

\begin{scope}[xshift=10.5cm,yshift=-2cm]
\draw (0,0) circle (0.5cm);
\node[draw=red] at (0,0)  {$\times$} ;
\end{scope}

\draw (-0.4,0) -- (9,-2);
\draw (0.4,0) -- (10.5,-2);

\begin{scope}[yshift=-2.7cm]
\draw [thick,decorate,decoration={brace,mirror,amplitude=10pt}] (2,0) -- (8.0,0);
\node[] at (5.0,-.6) {\large $p-1$}; 
\end{scope}

\end{tikzpicture}
\end{figure}

\noindent It gives rise to a new structure equal to $2 \times F^{p-1,0}_{p,0;4}$. Turning to Type II there are two relevant figures:
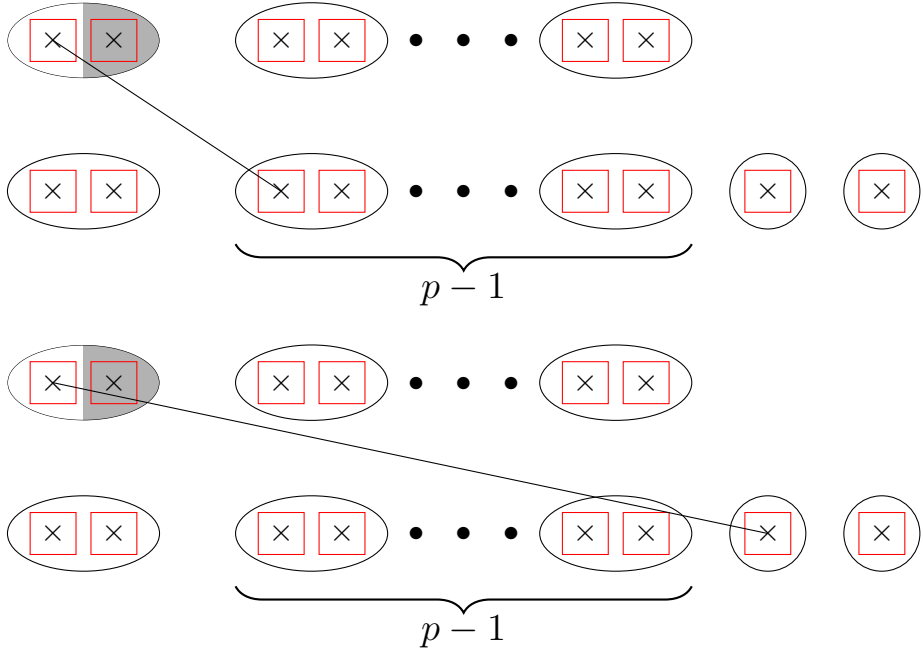
\begin{figure}[h]
\begin{tikzpicture}

\draw (5,-0.25) node[anchor=south] {\textbullet\quad\textbullet\quad\textbullet};

\begin{scope}
    \clip (0,0) ellipse (1cm and 0.5cm);
    \fill[black!30] (0cm,-1cm) rectangle (1cm, 1cm);
    \draw (0,0) ellipse (1cm and 0.5cm);
    \node[draw=red] at (-0.4,0)  {$\times$};
    \node[draw=red] at (0.4,0)  {$\times$} ;
\end{scope}

\begin{scope}[xshift=3cm]
\draw (0,0) ellipse (1cm and 0.5cm);
\node[draw=red] at (-0.4,0)  {$\times$};
\node[draw=red] at (0.4,0)  {$\times$} ;
\end{scope}
 
\begin{scope}[xshift=7cm]
\draw (0,0) ellipse (1cm and 0.5cm);
\node[draw=red] at (-0.4,0)  {$\times$};
\node[draw=red] at (0.4,0)  {$\times$} ;
\end{scope}

\draw (5,-2.25) node[anchor=south] {\textbullet\quad\textbullet\quad\textbullet};

\begin{scope}[yshift=-2cm]
\draw (0,0) ellipse (1cm and 0.5cm);
\node[draw=red] at (-0.4,0)  {$\times$};
\node[draw=red] at (0.4,0)  {$\times$} ;
\end{scope}

\begin{scope}[xshift=3cm,yshift=-2cm]
\draw (0,0) ellipse (1cm and 0.5cm);
\node[draw=red] at (-0.4,0)  {$\times$};
\node[draw=red] at (0.4,0)  {$\times$} ;
\end{scope}
 
\begin{scope}[xshift=7cm,yshift=-2cm]
\draw (0,0) ellipse (1cm and 0.5cm);
\node[draw=red] at (-0.4,0)  {$\times$};
\node[draw=red] at (0.4,0)  {$\times$} ;
\end{scope}

\begin{scope}[xshift=9cm,yshift=-2cm]
\draw (0,0) circle (0.5cm);
\node[draw=red] at (0,0)  {$\times$} ;
\end{scope}

\begin{scope}[xshift=10.5cm,yshift=-2cm]
\draw (0,0) circle (0.5cm);
\node[draw=red] at (0,0)  {$\times$} ;

\end{scope}
\draw (-0.4,0) -- (2.6,-2);

\begin{scope}[yshift=-2.7cm]
\draw [thick,decorate,decoration={brace,mirror,amplitude=10pt}] (2,0) -- (8.0,0);
\node[] at (5.0,-.6) {\large $p-1$}; 
\end{scope}

\end{tikzpicture}
\end{figure}

\begin{figure}[h]
\begin{tikzpicture}

\draw (5,-0.25) node[anchor=south] {\textbullet\quad\textbullet\quad\textbullet};

\begin{scope}
    \clip (0,0) ellipse (1cm and 0.5cm);
    \fill[black!30] (0cm,-1cm) rectangle (1cm, 1cm);
    \draw (0,0) ellipse (1cm and 0.5cm);
    \node[draw=red] at (-0.4,0)  {$\times$};
    \node[draw=red] at (0.4,0)  {$\times$} ;
\end{scope}

\begin{scope}[xshift=3cm]
\draw (0,0) ellipse (1cm and 0.5cm);
\node[draw=red] at (-0.4,0)  {$\times$};
\node[draw=red] at (0.4,0)  {$\times$} ;
\end{scope}
 
\begin{scope}[xshift=7cm]
\draw (0,0) ellipse (1cm and 0.5cm);
\node[draw=red] at (-0.4,0)  {$\times$};
\node[draw=red] at (0.4,0)  {$\times$} ;
\end{scope}

\draw (5,-2.25) node[anchor=south] {\textbullet\quad\textbullet\quad\textbullet};

\begin{scope}[yshift=-2cm]
\draw (0,0) ellipse (1cm and 0.5cm);
\node[draw=red] at (-0.4,0)  {$\times$};
\node[draw=red] at (0.4,0)  {$\times$} ;
\end{scope}

\begin{scope}[xshift=3cm,yshift=-2cm]
\draw (0,0) ellipse (1cm and 0.5cm);
\node[draw=red] at (-0.4,0)  {$\times$};
\node[draw=red] at (0.4,0)  {$\times$} ;
\end{scope}
 
\begin{scope}[xshift=7cm,yshift=-2cm]
\draw (0,0) ellipse (1cm and 0.5cm);
\node[draw=red] at (-0.4,0)  {$\times$};
\node[draw=red] at (0.4,0)  {$\times$} ;
\end{scope}

\begin{scope}[xshift=9cm,yshift=-2cm]
\draw (0,0) circle (0.5cm);
\node[draw=red] at (0,0)  {$\times$} ;
\end{scope}

\begin{scope}[xshift=10.5cm,yshift=-2cm]
\draw (0,0) circle (0.5cm);
\node[draw=red] at (0,0)  {$\times$} ;

\end{scope}
\draw (-0.4,0) -- (9.0,-2);

\begin{scope}[yshift=-2.7cm]
\draw [thick,decorate,decoration={brace,mirror,amplitude=10pt}] (2,0) -- (8.0,0);
\node[] at (5.0,-.6) {\large $p-1$}; 
\end{scope}

\end{tikzpicture}
\end{figure}

\noindent The first contributes $4 p \times F^{p-1,0}_{p-1,3;3}$ and the second $4 \times F^{p-1,0}_{p-1,1;3}$. Altogether we get the recursion relation
\bea
F^{p,0}_{p,2;4}=2 p (N+2 (p+1)) \times F^{p-1,0}_{p-1,2,4}+2 \times F^{p-1,0}_{p,0,4}+4 p \times F^{p-1,0}_{p-1,3;3}+4 \times F^{p-1,0}_{p-1,1;3}. \label{even2}
\eea

At this point, we have covered a fairly representative sample of the various kinds of contractions involved in the computations of this section. So now we will merely write down the rest of the recursion relations that are relevant in the determination of $F^{p,0}_{p,1;5}$, without belaboring the details. 
\bea
F^{p,0}_{p+1,1;3}=2(p+1)(2 (p+1)+N) \times F^{p-1,0}_{p,1;3} \\
F^{p,0}_{p,3;3}=2p (2(p+2)+N) \times F^{p-1,0}_{p-1,3;3}+6 \times  F^{p-1,0}_{p,1;3} \\
F^{p,0}_{p+1,0;4}=2(p+1)(N+2p) \times F^{p-1,0}_{p,0;4}+4 (p+1) \times F^{p-1,0}_{p-1,1;3}
\eea
These equations together with (\ref{even1},-\ref{even2}) together form a complete set of recursion relations which can be systematically solved for, once we provide the launching data at $p=0$. These are easily seen by inspection to be
\bea
F^{0,0}_{0,3;3}=1, \ \ F^{0,0}_{1,1;3}=1, \ \ F^{0,0,}_{1,0;4}=0, \ \ F^{0,0}_{0,2;4}=0, \ \ F^{0,0}_{0,1;5}=0.
\eea
With these initial conditions, the recursion relations can be trivially solved on Mathematica (we used the RecurrenceTable command) and the result is 
\bea
\rho_{2p}\equiv F^{p,0}_{p,1;5}/F^{p,0}_{p,1;1}=\frac{10 p^2 +(N-6) p}{2 (2+N)(4+N)}.
\eea
A nice consistency check of this result is that when we set $p=n/2$ and $N=1$ this expression reduces to $\rho=n(n-1)/12$ reproducing the results of the previous section.

A similar approach can be used to determine $\rho_{2p+1}$ as well, by starting with $F^{p,1}_{p+1,0;5}$. Again, we skip the details and present only the final complete set of recursion relations:
\bea
F^{p,1}_{p+1,0;5}=2(p+1)(N+2p)\times F^{p-1,1}_{p,0;5}+4(p+1) \times F^{p-1,1}_{p,1;4}+F^{p-1,1}_{p+1,0;3} \\
F^{p,1}_{p+1,1;4}=2(p+1)(N+2 (p+1)) \times F^{p-1,1}_{p,1;4}+4(p+1) \times F^{p-1,1}_{p,2;3}+2 \times F^{p-1,1}_{p+1,0;3} \\
F^{p,1}_{p+1,2;3}=2 (p+1) (N+2 (p+2)) \times F^{p-1,1}_{p,2;3} +2 \times F^{p-1,1}_{p+1,0;3} \\
F^{p,1}_{p+2,0;3}=2(p+2)(N+2(p+1))F^{p-1,1}_{p+1,0,3}.
\eea
Together with the initial conditions
\bea
F^{0,1}_{1,0;5}=0, \ \ F^{0,1}_{1,1;4}=1, \ \ F^{0,1}_{1,2;3}=4, \ \ F^{0,1}_{2,0;3}=4,
\eea
these can again be solved and the result is
\bea
\rho_{2p+1}\equiv F^{p,1}_{p+1,0;5}/F^{p,1}_{p+1,0;1}=\frac{10p^2+(3N+2)p}{2 (2+N)(4+N)}.
\eea
Again, it can be checked that for $p=(n-1)/2$ and $N=1$, this reduces to $\rho=n(n-1)/12$.

\subsection{Anomalous Dimensions}

Now we have all the ingredients necessary to set up the telescoping series and compute the anomalous dimensions along the lines of the previous section. The relevant $q$'s take the form
\bea
q_3^{2p}\approx -\frac{(\gamma_1+\gamma_{2p}-\gamma_{2p+1})}{12 \gamma_1}, \ \ q_3^{2p+1}\approx -\frac{(\gamma_1+\gamma_{2p+1}-\gamma_{2p+2})}{12 \gamma_1}
\eea
Demanding 
\bea
\lim_{\e \rightarrow 0}q_3^{i} \times \alpha = \rho_{i}, \ \ {\rm where} \ \ i=2p \ \ {\rm or}\  \ 2p+1,
\eea
together with (\ref{ONalpha}) forces 
\bea y_{1,1}=0 \eea and leads to the recursion relation
\bea
y_{i+1,1}-y_{i,1}=\sigma\rho_i \sqrt{96\ y_{1,2}(N+2)(N+4)} \equiv K_1\rho_i
\eea
where we have written the relations in terms of the Taylor series coefficients. In $d_0=3$ we further have $\Delta_5=\Delta_1+2$ which now becomes  $y_{5,1}=2$. This together with the recursion relations determines $\sigma=+1$ and 
\bea
y_{1,2}=\frac{(N+2)(N+4)}{24 (3N +22)^2}
\eea
which agrees with the result we found earlier for $N=1$. This also fixes $K_1$ to be $2(N+2)(N+4)/(3N+22)$. In terms of anomalous dimensions, we can write 

\bea
\gamma_{\phi^a}=\frac{(N+2)(N+4)}{24 (3N +22)^2}\e^2 + \mathcal{O}(\e^3) \label{Hagermatch}
\eea
We have checked that this result matches with perturbative loop computations, for example, in Hager \cite{Hager}, at two loop level\footnote{To make the comparison with Hager \cite{Hager}, we make a few comments about notation. We are using the Peskin{\&}Schroeder conventions for beta functions and anomalous dimensions. In particular, (19) in \cite{Hager} should be divided by two to match our anomalous dimension conventions. Moreover, (19) is written in terms of the coupling ($\bar{w}_R$ in \cite{Hager}), which we can solve in terms of $\e$ at the fixed point, by setting the beta function (18) to zero and solving for $\bar{w}_R$ at leading order. Plugging the resulting expression for $\bar{w}_R$ into (19) and dividing by the factor of two mentioned above, we find a precise match with (\ref{Hagermatch}).}. 

For completeness we also present the anomalous dimensions of general operators $W$ using our telescoping series:
\bea
y_{2p,1}=K_1 \left(\sum_{p'=0}^{p-1}\rho_{2p'+1}+\sum_{p'=1}^{p-1}\rho_{2p'}\right), \\
y_{2p+1,1}=K_1 \left(\sum_{p'=0}^{p-1}\rho_{2p'+1}+\sum_{p'=1}^{p}\rho_{2p'}\right),
\eea
Summing these expressions, we get the anomalous dimensions
\bea
\gamma_{\Phi_{2p}}=\frac{p(2p-2)(10p+3N-8)}{3 (22+3 N)} \e +\mathcal{O}(\e^2),\\
\gamma_{\Phi^a_{2p+1}}=\frac{p(2p-1)(10p+3N+2)}{3 (22+3 N)} \e +\mathcal{O}(\e^2),
\eea
both of which reduce (for even and odd $n$ respectively) to $\e n(n-1)(n-2)/30  +\mathcal{O}(\e^2)$ that we found in the previous section, when $N=1$.

\section{Comments on $d_0=6$ Theory}

Our discussion in the previous section was formally in generic $d_0$, but as we emphasized at various points, in practice there are restrictions arising from the fact that $r=2/(d_0-2)$ needs to be a positive integer. A example where this becomes evident is given by $d_0=6$ where the theory is a $\phi^3$ theory\footnote{After the first version of this paper, Yu Nakayama has informed us of some of his unpublished results in this direction which agree with our conclusions.}. The Lagrangian of the theory in dimension $d=6-\e$ is
\bea
S=\int d^{6-\epsilon} x \left(\frac{1}{2} \partial \phi^2 + \frac{g}{3!} \mu ^{\epsilon/2}\phi ^3 \right)
\eea
The multiplet shortening condition in this case is 
\bea
\Box \phi = \frac{g}{2!} \mu ^{\epsilon/2}\phi ^2
\eea
We can try to proceed as before to extract the $\e$-expansion from conformal field theory, by introducing Wilson-Fisher operators $V_n$ which tend to the free theory in the $\e \rightarrow 0$ limit. 

However there is one big difference in the flow of logic, which makes things different from before. This is because
\bea
\phi^n(x) \times \phi^{n+1}(0) \supset f^{n,n+1} |x|^{-4n}\{ \phi(0)+ ...\}
\eea
but the right hand side cannot contain $\phi^2$. We could also consider
\bea
\phi^n(x) \times \phi^{n+2}(0) \supset f^{n,n+2} |x|^{-4n}\{ \phi^2(0)\}
\eea
which does not have $\phi$ on the right hand side.
In these expressions, 
\bea
f^{n,n+1}=(n+1)!, \ \ f^{n,n+2}=(n+2)!/2!
\eea
It is clear that multiplet mixing in the naive sense that we used, is not going to be of immediate help here. 

These expressions imply that in the free theory limit (with $|x|\ll |z|$) \bea \langle V_n(x) V_{n+1}(0) V_1(z)\rangle \rightarrow \langle \phi^n(x) \phi^{n+1}(0) \phi(z)\rangle &\sim & f^{n,n+1} |x|^{-4n}\langle\phi(0)\phi(z) \rangle, \hspace{0.6in} \label{one}\\ \langle V_n(x) V_{n+1}(0) V_2(z)\rangle \rightarrow \langle \phi^n(x) \phi^{n+1}(0) \phi^2(z)\rangle & \sim & 0,   \label{two}\\ \langle V_n(x) V_{n+2}(0) V_1(z)\rangle \rightarrow \langle \phi^n(x) \phi^{n+2}(0) \phi(z)\rangle & \sim & 0,  \label{three} \\ \langle V_n(x) V_{n+2}(0) V_2(z)\rangle \rightarrow \langle \phi^n(x) \phi^{n+2}(0) \phi^2(z)\rangle &\sim & f^{n,n+2} |x|^{-4n}\langle\phi^2(0)\phi^2(z) \rangle, \hspace{0.6in} \label{four} \eea

One could try to look at how these limiting conditions constrain the coefficients in   \bea V_n(x) \times V_{n+1} (x) \supset \tilde f^{n,n+1} |x|^{\Delta_1-\Delta_n-\Delta_{n+1}}(1+ q^{n,n+1}_1 x^\mu \partial_\mu+ q^{n,n+1}_2 x^\mu x^\nu \partial_\mu\partial_\nu + q^{n,n+1}_3 x^2 \Box+...)V_1 (0) \nonumber \\ V_n(x) \times V_{n+2} (x) \supset \tilde f^{n,n+2} |x|^{\Delta_1-\Delta_n-\Delta_{n+2}}(1+ q^{n,n+2}_1 x^\mu \partial_\mu+ q^{n,n+2}_2 x^\mu x^\nu \partial_\mu\partial_\nu + q^{n,n+2}_3 x^2 \Box+...)V_1 (0) \nonumber \label{2ptexpn2} \eea

One can write down the expressions for the $q_3$'s as before but the condition that these expressions have a consistent free theory limit, does not immediately give any stringent requirements as it did before. In particular, we find that $y_{1,1}$ can have a contribution at $\mathcal{O}(\e)$, unlike in $d_0=3, 4$, and it is not determined by the arguments we have presented in the previous sections. The existence of this $\mathcal{O}(\e)$ term is consistent with the perturbative results of, eg., \cite{Bonfim}.

We will not explore this case further here, but this preliminary observation is enough to see why the case of $d_0=6$ is likely to have qualitative differences from the $d_0=3, 4$ cases.

\section*{Acknowledgments}

This note is a direct result of the the recently initiated ICTS-IISc joint seminar series: the authors became aware of \cite{Rychkov} because of the first seminar in this series. We thank the speaker, Kallol Sen, and the audience for a lively discussion.

\end{document}